# PENERAPAN TEKNOLOGI PENGOLAH CITRA DIGITAL DAN KOMPUTASI PADA PENGUKURAN DAN PENGUJIAN BERBAGAI PARAMETER KAIN RAJUT

Andrian Wijayono[1] & Valentinus Galih Vidia Putra[1]

Textile Engineering Departement, Politeknik STTT Bandung, Indonesia[1]

**Abstrak:** Salah satu tantangan besar industri saat ini adalah mengenai cara untuk memproduksi hasil produksi yang berkualitas, salah satunya adalah pada industri non woven. Peningkatan proses evaluasi dan pengendalian mutu produksi non woven telah banyak dikembangkan untuk mendukung peningkatan kualitas hasil produksi. Pemanfaatan teknologi informasi dan komputasi saat ini telah banyak diterapkan pada proses pengendalian mutu produksi bahan tekstil, salah satunya adalah pemanfaatan teknologi *image processing* pada proses evaluasi bahan kain rajut. Pada bab ini akan dijelaskan mengenai berbagai metoda penerapan teknologi *image processing* pada bidang evaluasi dan pengendalian mutu produksi tekstil.

**Kata Kunci:** *textile fiber, image processing, textile evaluation.*

# 1. PENDAHULUAN

Perkembangan teknik komputasi yang dinamis saat ini telah menciptakan kemungkinan besar untuk dapat diterapkan pada berbagai bidang, termasuk untuk mengidentifikasi dan mengukur dimensi geometris benda-benda yang sangat kecil termasuk benda tekstil. Dengan menggunakan analisis citra digital

# 2. SEGMENTASI POLA OTOMATIS PADA KAIN RAJUTAN LUSI JACQUARD BERDASARKAN METODE PENGOLAHAN CITRA HIBRID

Karena dapat menghasilkan berbagai variasi pola struktur pada kain, kain rajut lusi jacquard memegang peranan penting dalam kain rajutan, dan merupakan salah satu kainyang banyak digunakan pada industri garmen (Zhang dkk, 2015). Menurut Zhang dkk (2015), analisis pola secara manual dari sampel kain rajut lusi jacquard sangat membosankan dan memakan waktu, yang bisa memakan waktu hampir 70% dari total waktu proses untuk mendesain kain rajut. Oleh karena itu, diperlukan pengembangan metode analisis pola otomatis untuk sistem perancangan dengan komputer pada sistem rajut lusi.

Secara umum, literatur terkait pada studi analisis pola berfokus pada dua jenis kain. Beberapa studi difokuskan pada kain *printing*. Pada penelitian tersebut menggunakan algoritma genetika atau matriks co-occurrence digunakan untuk mengetahui nilai fitur pada citra, kemudian citra dianalisis menggunakan algoritma pemisahan warna, seperti *self-organization map* dan *neural network back-propagation* [Xu, 2002 dan Kuo, 2009]. Yang lainnya berkonsentrasi pada pemisahan warna dari kain bordir berdasarkan algoritma *clustering* GustafsonKessel (GK) [Kuo, 2012]. Namun, algoritma tersebut tidak dapat digunakan pada kain rajut lusi jacquard.

Pola kain rajutan lusi jacquard berasal dari struktur kain tersebut, yang terdiri dari *inlay*, *loop*, *fall plate* dan *lappet* (Zhang dkk, 2015). Sebagai contoh, terdapat tiga jenis jeratan rajutan lusi jacquard (Gambar 1), misalnya 1-0 / 1-2 // adalah tipe utama. Berdasarkan tipe utama, kita dapat memperoleh gerakan lainnya dengan pergeseran beberapa bilah dari gerakan asli, yaitu 1-0 / 2-3 //

dan 2-1 / 1-2 //. Permukaan kain 1-0 / 1-2 // adalah daerah seperti yang ditunjukkan pada Gambar-2a. Dengan perbandingan 1-0 / 1-2 //, underlap 2-1 / 1-2 // diperpendek, yaitu rajutan chaining dan daerah mesh pada permukaan kain seperti ditunjukkan pada Gambar-2b. Sebaliknya, underlap yang diperpanjang (Gambar-1b) akan mendapatkan tekstur permukaan yang tebal seperti ditunjukkan pada Gambar-2c. Dengan demikian, berbagai permukaan kain jacquard membentuk varietas tekstur yang berbeda ketika terkena cahaya.

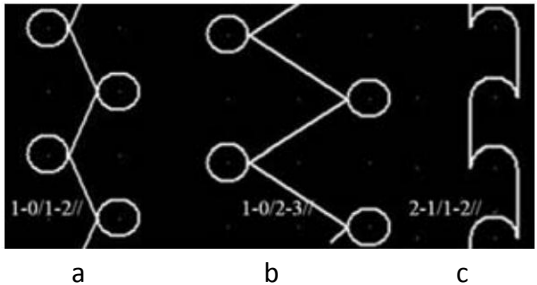

Gambar-1 Tiga jenis rajutan lusi jacquard, a) 1-0//1-2, b) 1-0//2-3 dan c) 2-1//1-2

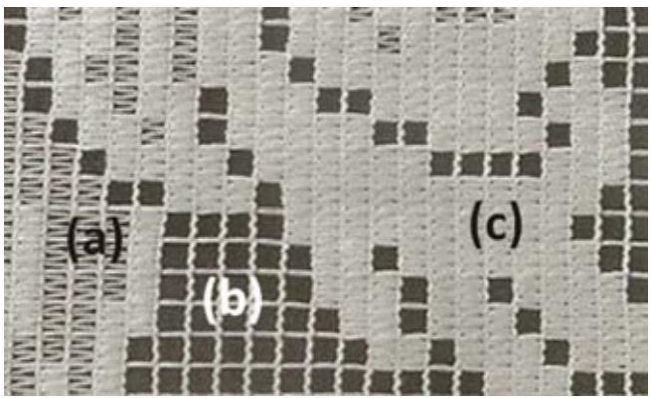

Gambar-2 Perbedaan kenampakan dari setiap jenis jeratan pada rajutan lusi jacquard, a) 1-0//1-2 , b) 2-1//1-2 dan c) 1-0//2-3

Mengingat karakteristik kain rajutan lusi jacquard, dalam penelitian Zhang dkk (2015) menerapkan metode pemisahan pola hybrid, termasuk *bilateral filter*, *pyramidal wavelet decomposition* dan *clustering* FCM yang telah diperbarui. Pertama, filter bilateral digunakan untuk mengurangi *noise* citra kain rajut lusi jacquard. Kedua, *pyramidal wavelet decomposition* dengan *wavelet Haar* digunakan untuk mengurangi beban perhitungan dan mempersingkat waktu perhitungan. Selanjutnya digunkaan metode peningkatan *clustering* dengan fungsi Mercer Kernel yang dapat menurunkan iterasi dan meningkatkan akurasi klasifikasi.

**Proses Akuisisi Citra Digital**

Pada penelitian penelitian Zhang dkk (2015), gambar kain rajutan lusi Jacquard ditangkap dan didigitasi oleh pemindai Canon CanoScan LiDE 210 yang dilengkapi dengan sensor gambar kontak (CIS), dioda pemancar cahaya tiga warna merah-hijau-biru (LED RGB) dan Universal Serial Bus hi-speed (USB) 2.0, dengan resolusi warna maksimum 4800 dpi × 4800 dpi dengan kedalaman warna 48-bit atau kedalaman grayscale 16 bit. Seperti yang ditunjukkan pada Gambar-3a, gambar tersebut adalah kain lingerie, dan gambar yang diperoleh telah dipindahkan ke yang terdiri dari skala abu-abu 8 bit, 512 pixel × 512 piksel dan disimpan dalam bitmap (BMP) format dengan 300 dpi.

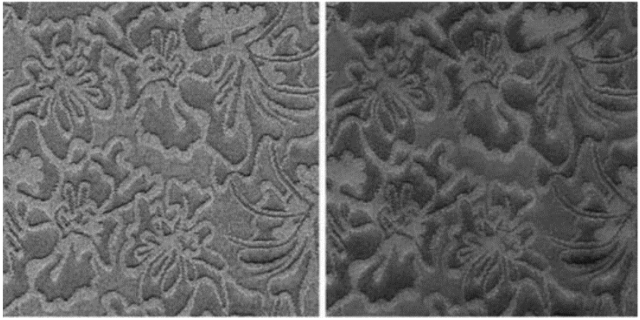

Gambar-3 Gambar kain rajut lusi jacquard, a) gambar asli dan b) gambar hasil filter

## Bilateral Filter

Berbagai gerakan *lapping* kain jacquard membentuk permukaan dan tekstur gambar tertentu. Permukaan kain tersebut akan menyebabkan refleksi yang berbeda di bawah iluminasi pemindai dalam proses pengambilan, yang menghasilkan dapat *noise*, seperti yang ditunjukkan di bawah ini

$$f(x,y) = f_o(x,y) + n(x,y) \qquad (1)$$

$f(x,y)$, $f_o(x,y)$ dan $n(x,y)$ masing-masing menunjukan gambar yang diambil, gambar asli dan noise pada citra kain.

Untuk mengurangi kebisingan dan menghaluskan tekstur secara bersamaan, Zhang dkk (2015) menggunakan filter bilateral. Penyaringan bilateral adalah teknik penyaringan non linier, yang mengembangkan konsep filter halus Gaussian dengan menimbang koefisien filter dengan intensitas piksel relatif [Lachkar, 2006]. Filter Gaussian dan bilateral didefinisikan di bawah ini

$$f(x,y) = \frac{1}{2\pi\sigma_s^2} e^{-\frac{x^2+y^2}{2\sigma_s^2}} \qquad (2)$$

$$\hat{f}(x,y) = \frac{\sum_{(i,j)\in S_{x,y}} w(i,j) g(i,j)}{\sum_{(i,j)\in S_{x,y}} w(i,j)} \qquad (3)$$

$$w(i,j) = e^{\frac{|i-x|^2 + |j-y|^2}{2\delta_s^2}} x e^{\frac{|g(i-j) - g(x,y)|^2}{2\delta_s^2}} \qquad (4)$$

## Pyramidal Wavelet Decomposition

Karena terdapat perbedaan refleksi pada permukaan kain yang berada di bawah iluminasi, sinyal gambar kain jacquard tidak bergerak, serta terletak baik dalam domain waktu dan frekuensi pada saat bersamaan. Oleh karena itu, Zhang dkk (2015) menggunakan transformasi *wavelet* yang berfungsi sebagai mikroskop matematika dalam pengolahan citra. Melalui transformasi *wavelet*, kita dapat mengamati dan menganalisa setiap bagian dari gambar kain dengan hanya menyesuaikan resolusi [Akay, 1995].

Algoritma dari *pyramidal wavelet decomposition* dapat dilihat sebagai berikut

$$f(x,y) \sum_{n=1}^{N} \sum_{k,l \in Z} (f_{HL}^{(n)} \psi_{n,k}(x)\phi_{n,l}(y) + f_{HL}^{(n)} \phi_{n,k}(x)\psi_{n,l}(y) \\ + f_{HH}^{(n)} \psi_{n,k}(x)\psi_{n,l}(y) + \sum_{k,l \in Z} f_{HH}^{(n)} \phi_{n,k}(x)\phi_{n,l}(y) \quad (5)$$

$$f_{LL}^{(n)} = \sum_{m,n,k,l \in Z} \bar{h}_{m-2k} \bar{h}_{m-2l} f_{LL}^{(n-1)} \quad (6)$$

$$f_{HL}^{(n)} = \sum_{m,n,k,l \in Z} \bar{g}_{m-2k} \bar{h}_{m-2l} f_{LL}^{(n-1)} \quad (7)$$

$$f_{LH}^{(n)} = \sum_{m,n,k,l \in Z} \bar{h}_{m-2k} \bar{g}_{m-2l} f_{LL}^{(n-1)} \quad (8)$$

$$f_{HH}^{(n)} = \sum_{m,n,k,l \in Z} \bar{g}_{m-2k} \bar{g}_{m-2l} f_{LL}^{(n-1)} \quad (9)$$

Hasil dari algoritma *pyramidal wavelet decomposition* dapat dilihat pada Gambar-4.

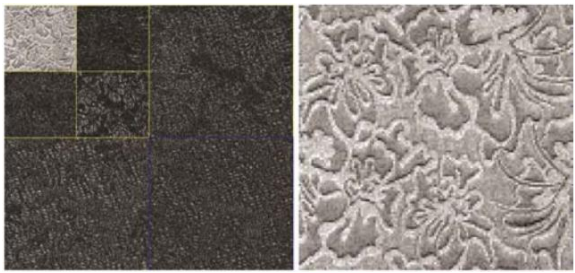

Gambar-4 Pengolahan *pyramidal wavelet decomposition* terhadap citra kain rajut lusi jacquard, a) *two scale wavelet decomposition* dan b) aproksimasi hasil algoritma *wavelet decomposition*

### *Improved* **FCM** *Clustering*

Terdapat dua metode *clustering*, yaitu *unsupervised clustering* dan *supervised clustering*. Untuk mengurangi waktu komputasi dan untuk memperbaiki efisiensi pemisahan pola, *unsupervised clustering* lebih sesuai untuk digunakan. Oleh karena itu, Zhang dkk (2015) menggunakan metode *clustering* FCM.

Pengelompokan FCM dikembangkan oleh Dunn dan Bezdek, yang memungkinkan satu data untuk dimiliki oleh dua atau lebih kategori fuzzy [Xie, 2009]. Namun, pengelompokan FCM tradisional bersifat sensitif terhadap data awal dan mudah terjebak dalam proses pengoptimalan local yang dilakukan oleh algoritma. Pada penelitian Zhang dkk (2015), telah digunakan *clustering* FCM terbarukan. Dalam metode tersebut, fungsi Mercer Kernel telah digunakan untuk mewujudkan peta dari ruang spesimen awal hingga ruang fitur berdimensi tinggi, yang membuat beberapa fitur menonjol untuk dilakukan proses pengelompokan (*clustering*).

$$J = \sum_{i=1}^{c}\sum_{k=1}^{n}(u_{ik})^m \|\emptyset(x_k) - \emptyset(v_i)\|^2 \tag{10}$$

$$\|\emptyset(x_k) - \emptyset(v_i)\|^2 = K(x_k, x_k) + K(v_i, v_i) - 2K(x_k, v_i) \tag{11}$$

$$K(x, y) = \emptyset(x)^T \emptyset(y) \tag{12}$$

$$J = \sum_{i=1}^{c}\sum_{k=1}^{n}(u_{ik})^m (1 - K(x_k, v_i)) \tag{13}$$

$$u_{ik} = \frac{\left(\frac{1}{(1-K(x_k,v_i))}\right)^{\frac{1}{m-1}}}{\sum_{j=1}^{c}\left(\frac{1}{(1-K(x_k,v_j))}\right)^{\frac{1}{m-1}}} \tag{14}$$

$$v_i = \frac{\sum_{k-1}^{n}\mu_{ik}^m K(x_k, v_j) x_k}{\sum_{k-1}^{n}\mu_{ik}^m K(x_k, v_j)} \tag{15}$$

$$x_{kj} = \frac{\sum_{i=1}^{c}\mu_{ik}^m K(x_k, v_j) v_i}{\sum_{i=1}^{c}\mu_{ik}^m K(x_k, i)} \tag{16}$$

Pendekatan yang diusulkan oleh Zhang dkk (2015) telah dilakukan pada platform MATLAB R2011a dengan PC yang dilengkapi dengan prosesor Intel Core i3-3240 (3M Cache, 3.4GHz) dan SDRAM DDR3 Dual Channel 4GB pada 1600MHz. Untuk mengevaluasi kinerja metode yang diajukan, dua indeks terukur digunakan, yaitu koefisien Kapp (KC) dan klasifikasi accuracy ratio (CAR) [Xiao, 2012]. KC adalah ukuran statistik perjanjian antar-penilai atau perjanjian antar-annotator untuk item kategoris. CAR menunjukkan persentase kategorisasi pixel yang benar. Algoritma yang digunakan dapat dilihat sebagai berikut

$$KC = \frac{N \sum N_{ii} - \sum(N_{i+}N_{+i})}{N^2 - \sum(N_{i+}N_{+i})} x 100\% \quad (17)$$

$$CAR = \frac{\sum_{i=1}^{M} N_{ii}}{N} x 100\% \quad (18)$$

$$N_{i+} = \sum_{j=1}^{M} N_{ij} \quad (19)$$

$$N_{+i} = \sum_{i=1}^{M} N_{ji} \quad (20)$$

dimana $N$ dan $M$ mewakili jumlah piksel dan kelompok, dan $N_{ii}$ menunjukkan jumlah klasifikasi akurat dalam cluster $i$.

Terlepas dari pendekatan yang diusulkan, ada dua metode FCM tradisional yang digunakan untuk perbandingan, yaitu FCM 1# dan FCM 2#. FCM 1 # menggunakan filter bilateral, dekomposisi multi-skala dan pengelompokan FCM dengan fungsi kernel tradisional. FCM 2# menggunakan filter Gaussian dan FCM clustering dengan fungsi kernel tradisional.

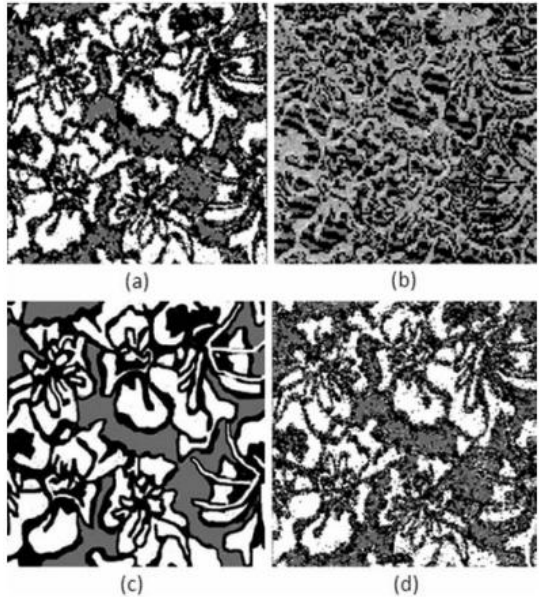

Gambar-5 Hasil operasi *pattern separation,* a) dengan FCM 1#, b) dengan FCM 2#, c) dengan metode yang diusulkan Zhang dkk (2015) dan d) metode MRF

Seperti ditunjukkan pada Gambar-5 dan indeks kuantitatif terkait pada Tabel-1, di bawah prasyarat yang sama, iterasi praktis FCM 1# memiliki lebih dari dua kali lipat dari metode yang diajukan Zhang dkk (2015), yang disertai dengan waktu komputasi yang lebih lama dan ketepatan klasifikasi yang lebih rendah. Perbedaan antara FCM 1# dan metode yang kami ajukan menegaskan fakta bahwa metode pengelompokan FCM tradisional sangat peka terhadap *noise* dan tidak dapat mengatasi masalah optimasi lokal (Gambar-5a). Namun sebaliknya, fungsi Mercer Kernel dengan bobot serupa dalam metode yang diusulkan dapat menurunkan iterasi dan meningkatkan akurasi klasifikasi. Kinerja FCM 2# tidak dapat diterima. Pertama, filter Gaussian menghaluskan tekstur permukaan jacquard, namun tidak dapat mempertahankan tepi gambar (Gambar-5b). Kemudian waktu komputasi FCM 2# menjadi empat kali lipat dari metode yang diajukan, karena algoritma tanpa transformasi *wavelet* akan meningkatkan beban perhitungan dan waktu komputasi untuk *clustering* berikutnya.

Selain mengevaluasi karakteristik ini, peneliti masih membandingkan metode yang diusulkan dengan pendekatan segmentasi tekstur lainnya, seperti bidang acak Markov (MRF) yang merupakan salah satu algoritma grafis probabilistik yang khas. Meskipun metode MRF tidak didedikasikan untuk gambar kain jacquard. Algoritma tersebut berperilaku baik dan memiliki beragam aplikasi dalam segmentasi tekstur. Seperti ditunjukkan pada Gambar-5d dan informasi yang sesuai pada Tabel-1, metode MRF rentan terhadap *noise*, yang akan sangat meningkatkan iterasi dan mudah untuk beralih ke pengoptimalan lokal.

Tabel-1 Indeks kuantitatif dari hasil operasi *pattern separation*

| Metode | Jumlah *clustering* | *Terminal condition* | Iterasi maksimum | Iterasi yang digunakan | Waktu komputasi | KC | CAR |
|---|---|---|---|---|---|---|---|
| FCM 1# | 3 | 0.01 | 100 | 20 | 17 | 83.8 | 86.3 |
| FCM 2# | 3 | 0.01 | 100 | 38 | 32 | 45.2 | 51.6 |
| Yang diajukan | 3 | 0.01 | 100 | 8 | 8 | 95.32 | 97.25 |
| MRF | 3 | / | / | / | 58 | 75.1 | 78.9 |

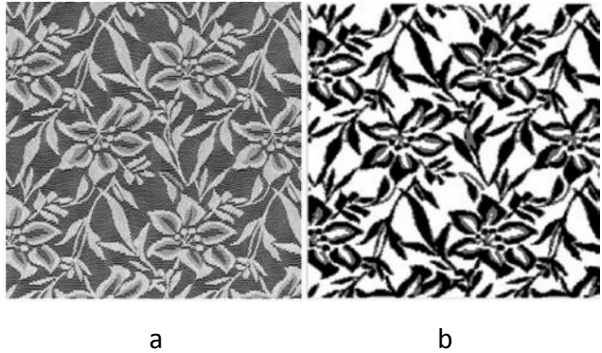

<p style="text-align:center">a          b</p>

Gambar-6 Hasil operasi *pattern separation*, a) sebelum dan b) sesudah

Oleh karena itu, metode pemisahan pola hibrid termasuk *bilateral filter*, *pyramidal wavelet decomposition* dan *clustering* FCM yang lebih baik cocok untuk pemisahan pola kain rajutan lusi jacquard, seperti yang ditunjukkan pada Gambar-5c dan Gambar-6.

Filter bilateral digunakan untuk menghaluskan tekstur kain yang dibentuk oleh berbagai gerakan *lapping* kain jacquard dan untuk mengurangi *noise* yang muncul dalam proses pengambilan citra, namun juga dapat menjaga tepi gambar. *Pyramidal wavelet decomposition* dengan basis *wavelet* Haar digunakan untuk memproses sinyal kain jacquard non-stasioner, perkiraan sub-gambar menyimpan sebagian besar informasi dan data spesifik gambar asli. Oleh karena itu, proses *clustering* selanjutnya harus dapat mengurangi beban perhitungan dan mempersingkat waktu komputasi. Mengingat pengelompokan FCM tradisional sensitif terhadap *noise* dan mudah dijebak ke pengoptimalan lokal, Zhang dkk (2015) mengusulkan penggabungan FCM yang dimodifikasi, dengan tujuan menonjolkan beberapa fitur dengan mengelompokkan fungsi Mercer Kernel yang digunakan untuk mewujudkan peta dari ruang spesimen awal ke ruang fitur dimensi tinggi, dan fungsi bobot diusulkan untuk mengukur kemiripan antara data dan pusat pengelompokan. Hasil penelitian menunjukkan bahwa metode pemisahan pola hibrida layak dilakukan dan dapat diterapkan.

# 3. TINJAUAN PERILAKU BAGGING KAIN RAJUT DENGAN BERBAGAI STRUKTUR DAN BENANG CAMPURAN MENGGUNAKAN IMAGE PROCESSING

Pencampuran serat (*fiber blending*) umumnya didefinisikan sebagai proses pembentukan benang dari campuran komponen serat yang berbeda (Hasani dkk, 2012). Benang campuran dari serat alami dan buatan memiliki beberapa kelebihan, khususnya untuk menggabungkan sifat baik yang tidak dapat ditemukan hanya pada satu jenis serat. Benang campuran serat viskosa/poliester biasanya diproduksi di industri tekstil karena beberapa keunggulan seperti *pilling* yang lebih sedikit, elektrostatik yang lebih sedikit, lebih mudah dipintal dan memiliki kerataan benang yang lebih baik [Baykal, 2007].

*Bagging* didefinisikan sebagai deformasi residual tiga dimensi, biasanya terlihat pada pakaian bekas, yang menyebabkan berkurangnya nilai penampilan kain. Tempat yang terlihat saat dipakai adalah siku, lutut, saku, pinggul, dan tumit [Amirbayat, 2005]. Bagging diakibatkan oleh kurangnya stabilitas dimensi atau pemulihan saat tekanan berulang atau berkepanjangan diberikan pada kain [Amirbayat, 205].

Untuk mengevaluasi perilaku *bagging*, beberapa metode untuk menentukan perilaku *bagging* kain tenun dan rajutan telah dikembangkan [Jaoachi, 2010, Joudnukyte, 2006, Ozdil, 2008, Zhang, 1998, Zhang, 1999, Yeung, 2002]. Sebagian besar publikasi berfokus pada pengukuran tinggi *bagging* dan sifat mekanik kain terkait. Yokura dkk telah mengukur sifat mekanik dari kain menggunakan sistem KES-FB untuk memprediksi volume *bagging* dari kain tenunan dari sifat kain yang diukur. Zhang dkk (1999, 1998 dan 2000) juga mengukur tinggi *bagging* kain tenunan dengan alat penguji tarik Instron. Sampel kain ditarik dalam lima siklus beban, tinggi *bagging* dan ketahanan *bagging* diukur. Zhang dkk menggunakan analisis regresi untuk memprediksi tinggi *bagging* kain tenun sebagai fungsi ketahanan. Karena perbedaan struktural, respon mekanis kain rajutan sangat berbeda dengan kain tenunan. Dengan demikian perilaku *bagging* kain rajutan berbeda dengan kain tenunan.

Uçar dkk [2002] membahas hubungan antara tinggi *bagging* yang diperoleh dari uji pengacakan kain dan karakterisasi mekanis yang ditentukan dari sistem KES-FB. Uçar dkk [2002] memperkirakan tinggi pengantongan residu untuk kain rajutan dengan menggunakan uji KESFB standar. Yeung dan Zhang [2002] mengembangkan metode untuk mengevaluasi *bagging* pakaian dengan pengolahan gambar dengan teknik pemodelan yang berbeda.

Hasani dkk (2012) telah meneliti efek rasio campuran dan struktur kain pada berat *bagging* kain rajutan yang dihasilkan dari benang rotor campuran viscose / poliester dengan teknik analisis citra. Desain kisi simpleks digunakan untuk menentukan kombinasi rasio campuran jenis serat.

Serat viskosa dan poliester dicampur dan dipintal pada sistem pemintalan rotor. Serat yang diproses pada sistem ini menggunakan prosedur, penyesuaian dan praktik standar pabrik. Sliver material viskosa dicampur dengan serat poliester pada mesin *drawing breaker,* kemudian *sliver* dicampur dan dilewatkan melalui *drawing passage* dua. Potongan viscose / polyester dicampur dan digunakan untuk menghasilkan benang Ne$_1$30 pada mesin pemintalan rotor pada kondisi atmosfir standar. Spesifikasi dari benang yang dihasilkan ditunjukkan pada Tabel-2. Menggunakan mesin *double jersey*, *mini-jacquard*, mesin rajut bundar (Mayer & Cie, E20, 30 ") yang dilengkapi dengan mekanisme *feeding* positif, tiga struktur rajutan dibuat: yaitu (1) *plain interlock*, (2) *tuck cross interlock* (yang terdiri dari *tuck* dan *loop stitch*) dan *miss cross interlock* (yang terdiri dari *miss* dan *loop stitch* pada mesin rajut bundar). Masing-masing kain rajutan mengandung rasio perpaduan serat yang berbeda dan desain kain yang berbeda sehingga kita bisa menyelidiki pengaruh desain dan bahan kain (tipe serat) terhadap perilaku *bagging* kain. Diagram jeratan masing-masing sampel dapat dilihat pada Gambar-7. Sampel dikondisikan selama 24 jam dalam atmosfir standar. Wale dan course count per 100 cm kain diukur dan kemudian dikonversi menjadi wale dan course count per cm. Panjang stitch dari kain rajutan diukur untuk menentukan *stitch length*.

Tabel-2 Spesifikasi benang yang dibuat dengan rasio campuran yang berbeda

| Design Points | Blend ratios | | Yarn count (Ne) | Yarn twist (tpm) |
|---|---|---|---|---|
| | Viscose | Polyester | | |
| 1 | 0 % | 100 % | 30.11 | 640 |
| 2 | 25 % | 75 % | 29.90 | 635 |
| 3 | 50 % | 50 % | 29.94 | 618 |
| 4 | 75 % | 25 % | 30.10 | 660 |
| 5 | 100 % | 0 % | 29.79 | 652 |

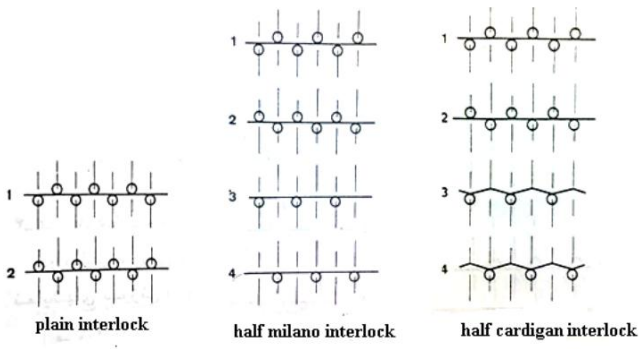

Gambar-7 Diagram jeratan tiap sampel

Untuk mempersiapkan sampel dalam keadaan relaksasi basah, kain dicuci di mesin cuci pada suhu 40°C selama 30 menit dengan deterjen komersial dan dikeringkan pada suhu 70°C selama 15 menit. Prosedur tersebut diulang sebanyak tiga kali. Sampel dikondisikan selama 24 jam dalam atmosfir standar. Desain kisi simpleks dengan tujuh ulangan pada setiap titik disain dibuat untuk menentukan kombinasi rasio campuran dari dua jenis serat. Dalam penelitian ini, desain kisi simpleks {2, 4} digunakan untuk menentukan campuran viscose / polyester. Poin desain (rasio campuran) yang digunakan dalam penelitian ini ditunjukkan pada Tabel-2. Sebelum pengukuran dilakukan, *stitch length* dari rata-rata sepuluh pengukuran dari masing-masing sampel digunakan dalam persamaan berikut untuk mendapatkan faktor ketat (tightness factor / TF) dari kain rajutan *double jersey*

$$T.F = \frac{\sqrt{Tex}.N_c}{l_c} \qquad (21)$$

$l_c$ merupakan nilai *stitch length* pada kain rajut dalam satu repeat struktur, sedangkan $N_c$ adalah jumlah jarum yang aktif dalam satu repeat struktur kain.

Berat kain diperoleh dari rata-rata tiga pengukuran masing-masing sampel dengan menggunakan timbangan, dan dilaporkan dalam g/m2. Berat kain, kerapatan rajutan dan faktor keketatan yang dihitung dari masing-masing sampel ditunjukkan pada Tabel-3. Perbedaan antara faktor ketat masing-masing struktur rajutan yang dihasilkan dari rasio perpaduan yang berbeda tidak signifikan secara statistik. Jadi rata-rata nilai ini dihitung untuk setiap struktur kain dan dilaporkan pada Tabel-4.

Tabel-3 Spesifikasi kain rajut yang digunakan

| Property | Blend ratio | | Weight (gr/m²) | Thickness (mm) | SD (loop/cm²) | T.F |
|---|---|---|---|---|---|---|
| | Polyester | Viscose | | | | |
| Plain interlock (S1) | 1.000 | 0.000 | 244.12 | 0.812 | 183.87 | 24.12 |
| | 0.000 | 1.000 | 242.56 | 0.809 | 185.89 | 23.21 |
| | 0.750 | 0.250 | 241.67 | 0.809 | 178.98 | 24.98 |
| | 0.500 | 0.500 | 242.98 | 0.804 | 180.67 | 23.89 |
| | 0.000 | 1.000 | 243.98 | 0.812 | 184.89 | 24.09 |
| Half milano interlock (S2) | 1.000 | 0.000 | 91.09 | 0.902 | 89.10 | 20.89 |
| | 0.000 | 1.000 | 90.78 | 0.877 | 90.12 | 23.21 |
| | 0.750 | 0.250 | 89.98 | 0.910 | 88.88 | 22.98 |
| | 0.500 | 0.500 | 87.98 | 0.887 | 92.12 | 21.79 |
| | 0.000 | 1.000 | 90.54 | 0.911 | 89.99 | 21.43 |
| Half cardigan Interlock (S3) | 1.000 | 0.000 | 104.56 | 0.997 | 104.02 | 28.3 |
| | 0.000 | 1.000 | 104.87 | 0.978 | 101.45 | 27.01 |
| | 0.750 | 0.250 | 101.64 | 1.002 | 102.87 | 26.56 |
| | 0.500 | 0.500 | 102.87 | 0.998 | 103.44 | 27.12 |
| | 0.000 | 1.000 | 102.44 | 0.998 | 104.87 | 25.96 |

Tabel-4 Hasil pengujian *tensile, shear* dan *drape*

| Structure | Drape Coef. | WT[1] | RT[2] | G[3] | 2HG[4] |
|---|---|---|---|---|---|
| S1 | 0.75 | 11.4 | 0.310 | 0.189 | 0.53 |
| S2 | 0.79 | 9.4 | 0.257 | 0.193 | 0.72 |
| S3 | 0.66 | 7.5 | 0.188 | 0.174 | 0.32 |

Prosedur untuk mengevaluasi pengepakan kain meliputi pengambilan gambar digital dari *bagging* kain, pemrosesan gambar yang diambil, pemilihan kriteria untuk menggambarkan penampilan *bagging*, dan mengetahui besarnya *bagging* dari kriteria ini. Lima belas kain rajut diuji dengan menggunakan metode uji *bagging* yang dikembangkan sebelumnya. Pada waktu yang telah ditentukan setelah kain itu dikantongi, mereka dipotret dengan kamera CCD dan disimpan sebagai file digital. Dalam proses pengambilan foto, semua gambar ditransfer ke dalam bentuk gambar intensitas. Intensitas gambar mengacu pada fungsi intensitas cahaya dua dimensi, dilambangkan dengan $f(x, y)$. Pada gambar digital, nilai intensitas pada koordinat $(x, y)$ atau tingkat abu-abu pada titik itu terletak pada kisaran $(0, 255)$, 0 untuk hitam dan 255 untuk warna putih. Parameter teknis, seperti pembesaran, posisi, kecerahan, dan sudut sumber cahaya, dijaga tetap sama. Gambar yang diambil dianalisis dengan menggunakan perangkat lunak Matlab. Analisis ini menghasilkan kurva simulasi *bagging*. Gambar-8 menunjukkan kain yang diuji *bagging* dan kurva simulasi *bagging* yang dicapai dengan analisis citra. Tingginya ketinggian *bagging* tujuh sampel dihitung menurut Persamaan (21) dan rata-rata pengukuran dilaporkan. *Bagging residual* tinggi kain rajutan yang diukur dengan menggunakan teknik pengolahan citra ditunjukkan pada Tabel-5.

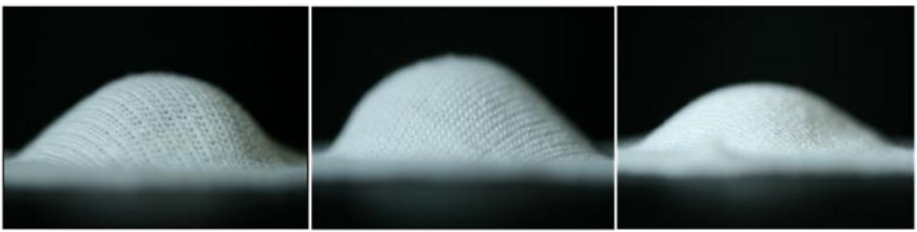

(a)           (b)           (c)

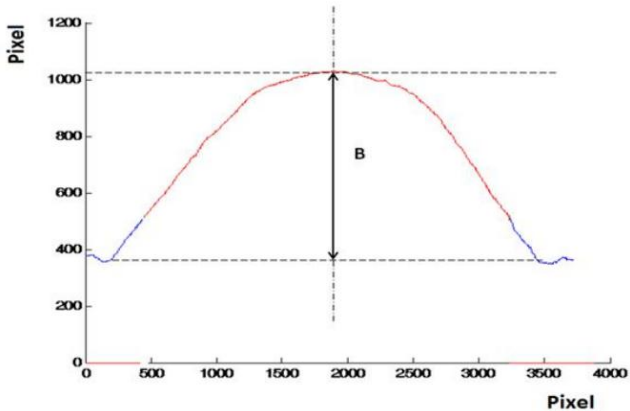

(d)

Gambar-8 Hasil pengujian *bagging* pada struktur rajut a) *half cardigan interlock,* b) *half milano interlock.* c) *plain interlock* dan d) merupakan contoh kurva *bagging* yang dianalisis menggunakan metode pengolahan citra digital

Tabel-5 Nilai persentase ketinggian *residual bagging* dari berbagai struktur kain rajut

| Run | Polyester ratio (A) | Viscose ratio (B) | Structure | T.F | Residual bagging height (%) |
|---|---|---|---|---|---|
| 1 | 1.000 | 0.000 | Half cardigan interlock | 26.97 | 72.97 |
| 2 | 0.000 | 1.000 | Plain interlock | 22.05 | 69.80 |
| 3 | 0.750 | 0.250 | Plain interlock | 22.05 | 65.63 |
| 4 | 0.500 | 0.500 | Half cardigan interlock | 26.97 | 75.76 |
| 5 | 0.000 | 1.000 | Half milano interlock | 24.06 | 79.18 |
| 6 | 0.250 | 0.750 | Half milano interlock | 24.06 | 77.79 |
| 7 | 0.750 | 0.250 | Half milano interlock | 24.06 | 75.01 |
| 8 | 0.250 | 0.750 | Half cardigan interlock | 26.97 | 77.15 |
| 9 | 0.500 | 0.500 | Half milano interlock | 24.06 | 76.34 |
| 10 | 0.000 | 1.000 | Half cardigan interlock | 26.97 | 78.55 |
| 11 | 0.750 | 0.250 | Half cardigan interlock | 26.97 | 74.37 |
| 12 | 1.000 | 0.000 | Plain interlock | 22.05 | 64.24 |
| 13 | 0.250 | 0.750 | Plain interlock | 22.05 | 68.42 |
| 14 | 1.000 | 0.000 | Half milano interlock | 24.06 | 73.61 |
| 15 | 0.500 | 0.500 | Plain interlock | 22.05 | 67.02 |

Hasil menunjukkan bahwa peningkatan persentase viscose akan meningkatkan tinggi *residual bagging* kain. Fenomena ini diamati pada struktur kain yang berbeda. Gambar-9 menunjukkan tinggi *residual bagging* yang terkait dengan persentase viskosa pada benang untuk struktur kain yang berbeda. Dua penyebab utama dari perilaku *bagging* kain adalah sifat relaksasi stres dari serat, karena perilaku viskoelastis serat, gesekan antara serat dan benang, dan hambatan friksi pada struktur kain. Sifat mekanik serat-benang dan sifat struktural kain, seperti ketebalan kain, berat, faktor keteguhan dan titik interlacing, adalah faktor penting yang mempengaruhi perilaku *bagging* kain [Kirk dan Ibrahim, 1966]. Temuan menunjukkan bahwa seiring dengan pertambahan kandungan persentase serat viscose dalam benang, menyebabkan nilai tinggi *bagging residual* kain meningkat. Untuk serat poliester, rasio elastisitasnya tinggi dan rasio viskoelastisitasnya rendah. Untuk serat viscose, rasio elastisitasnya rendah dan rasio viskoelastisitasnya tinggi. Selain itu, waktu relaksasi untuk serat poliester lebih tinggi dari serat viscose [Sengoz, 2004]. Model regresi yang paling tepat yang mendefinisikan hubungan antara variabel independen (rasio campuran dan faktor keketatan kain) dan variabel respon (tinggi pengantung residu) dipilih dan diperkirakan menggunakan perangkat lunak *Design Expert Software*. Pengambilan kain rajutan viscose / polyester dapat diprediksi untuk rasio pencampuran yang berbeda dan faktor keketatan kain dengan menggunakan persamaan berikut

$$B_{residual}(\%) = 69.5 - 5{,}56\,P + 0{,}31 T.F + 9{,}32 T.F^2 \qquad (22)$$

Dalam persamaan ini, "P" adalah kandungan poliester dari benang campuran dan T.F adalah faktor keteguhan dari kain rajutan. Gambar-9 mengilustrasikan kurva regresi yang sesuai dengan pengamatan eksperimental. Koefisien korelasi yang didapatkan adalah sebesar 0,999, menunjukkan kemampuan prediktif kuat dari model regresi yang dibangun. Tabel ANOVA untuk model regresi dan koefisien estimasinya ditunjukkan pada Tabel-6. Nilai F Model menunjukkan bahwa model tersebut signifikan secara statistik.

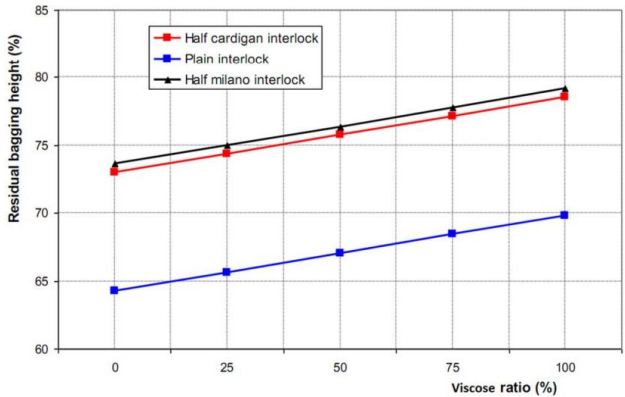

Gambar-9 Hubungan antara ketinggian *residual bagging* dan kandungan serat viskosa pada kain untuk jenis struktur kain yang berbeda

Tabel-5 Hasil analisis ANOVA

| Source | SS | Df | MS | F value | P-value |
|---|---|---|---|---|---|
| Model | 332.21 | 5 | 66.44 | 2.143E+005 | 0.0001 |
| Linear mixture | 58.16 | 1 | 58.16 | 1.876E+005 | 0.0001 |
| AC | 0.33 | 1 | 0.33 | 1066.94 | 0.0001 |
| BC | 0.32 | 1 | 0.32 | 1040.01 | 0.0001 |
| $AC^2$ | 88.76 | 1 | 88.76 | 2.863E+005 | 0.0001 |
| $BC^2$ | 88.93 | 1 | 8.93 | 2.869E+005 | 0.0001 |
| Residual | 2.790E-003 | 9 | 3.100E-004 | | |
| Total | 332.21 | 14 | | | |

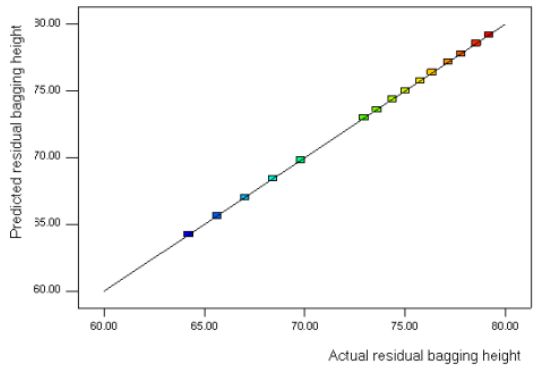

Gambar-10 Hubungan regresi antara prediksi dan eksperimen

Berdasarkan penelitian yang telah dilakukan oleh Hasani dkk (2012), tinggi *bagging residual* kain rajutan viscose-polyester dimodelkan melalui model regresi dimana rasio campuran dan struktur kain adalah variabel prediktor. Model ini memiliki kemampuan prediksi tinggi yang ditunjukkan oleh korelasi positif tinggi antara nilai tinggi pengantung residu yang diperkirakan dan nilai tinggi pengangkutan yang diamati. Apabila persentase serat viscose dalam campuran meningkat, maka tinggi *bagging residual* juga ikut meningkat. Hal ini dapat disebabkan oleh modulus viskoelastis yang lebih tinggi dan waktu hubungan serat viscose yang lebih kecil. Juga, temuan tersebut mengungkapkan bahwa tinggi kain *bagging residual* lebih rendah pada struktur yang dihasilkan dari jahitan yang tidak rata. Kain rajutan dengan struktur interlock polos yang dihasilkan dari benang poliester 100% memiliki *bagging residual* terendah

## 4. EVALUASI STRUKTUR KAIN hASCs RAJUT 3D DENGAN MENGGUNAKAN ANALISIS CITRA

Di bidang biomedis, teknik jaringan (*tissue engginering* disingkat TE) mewakili area spesifik di mana teknologi tekstil dapat memiliki kontribusi penting [Mather, 2006 dan Wollina, 2003]. Memaksimalkan keterikatan jaringan pada bahan memerlukan struktur berpori yang sangat teratur untuk integrasi jaringan dan kerangka untuk perakitan sel, dikombinasikan dengan sifat struktural yang serupa dengan jaringan kehidupan. Perancah dikembangkan untuk TE, perlu memfasilitasi dan mempromosikan proliferasi seluler dan regenerasi jaringan. Geometri perancah intraarchitect, serta bahan perancah dan sifat permukaannya memainkan peran penting dalam proses ini. Banyak teknik fabrikasi konvensional yang tersedia untuk produksi perancah belum memungkinkan untuk mendapatkan properti perancah yang diinginkan, karena banyak tahap pemrosesan yang masih memiliki kemampuan reproduktifitas yang lambat. Beberapa metode telah dikembangkan dan diusulkan untuk menyiapkan perancah berpori untuk rekayasa jaringan, termasuk pembentuk gas, ekstrusi serat dan ikatan, pencetakan tiga dimensi, pemisahan fasa, pengeringan beku emulsi, dan pelindian porogen atau prototipe cepat [Hutmatcher, 2001 dan Salgado, 2004]. Sebagian besar teknik ini telah dipelajari secara ekstensif dengan menggunakan berbagai polimer biodegradable, seperti asam poliglikolat (PGA), asam polylactic acid (PLA), polycrapolactone (PCL), pati atau sutra, yang diselidiki untuk transplantasi sel dan regenerasi berbagai jaringan, seperti saraf, kulit, ligamen, kandung kemih, tulang rawan, dan tulang [Salgado dkk, 2004, Burg dkk, 2000, Gomes dkk, 2008]. Perancah yang diproses dengan menggunakan teknologi berbasis serat mewakili berbagai kemungkinan morfologi dan geometrik yang dapat disesuaikan untuk setiap aplikasi teknik jaringan tertentu [Tuzlakoglu, 2009]. Sebenarnya, jaringan serat dengan luas permukaan yang tinggi dan interkonektivitas telah terbukti sangat mendukung pelekatan dan proliferasi sel [Silva dkk 2010, Gomes dkk, 2008]. Oleh karena itu, teknologi berbasis tekstil dianggap sebagai langkah potensial untuk produksi perancah kompleks untuk aplikasi TE, karena dapat memberikan kontrol yang superior atas desain, tingkat presisi manufaktur dan reproduktifitas yang baik.

Tujuan dari penelitian yang dilakukan Ribeiro dkk (2013) adalah untuk mengevaluasi potensi struktur biotekstil yang baru untuk dikembangkan sebagai perancah untuk rekayasa jaringan [Almeida, 2011 dan Ribeiro, 2011]. Polybuthylene suksinin (PBS) pada awalnya diusulkan sebagai serat multifilamen ekstrusi yang dapat diproses dengan teknologi berbasis tekstil. Sebuah studi komparatif dibuat menggunakan serat SF dengan kerapatan linier serupa. Teknologi rajut digunakan untuk membuat matriks tekstil biodegradable. Alasan rasional untuk menggunakan teknologi ini adalah bahwa substrat tekstil rajutan diketahui menunjukkan kemampuan atau kesesuaian yang lebih baik dibandingkan dengan substrat tenunan lainnya, dengan porositas / volume yang disempurnakan meskipun dengan ketebalan yang terbatas [Wang, 2011]. Untuk mengatasi hal ini, benang sutera alam juga diproses untuk pertama kalinya ke struktur 3D yang berbeda dengan menggunakan teknologi rajut lusi untuk meningkatkan tridimensionalitas perancah. Dalam kasus terakhir, dua lapisan sutra rajutan dirakit dan diberi jarak oleh monofilamen polietilen tereftalat (PET). Setiap jenis serat polimer dapat memungkinkan pembangkitan konstruksi dengan karakteristik yang berbeda dalam hal fisik permukaan, kemampuan mekanik dan kemampuan degradasi, yang berdampak pada perilaku sel yang dihasilkan pada permukaan biotextiles masing-masing. Skrining sitotoksisitas awal menunjukkan bahwa kedua bahan tersebut dapat mendukung adhesi sel dan proliferasi setelah diolah. Selanjutnya, modifikasi permukaan yang berbeda dilakukan (perawatan asam / basa, radiasi UV dan plasma) untuk memodulasi perilaku sel. Human Adipose derived Stem Cells (hASCs) menjadi kemungkinan munculnya terapi penggantian jaringan. Potensi struktur biotekstil sutra yang baru dikembangkan untuk mempromosikan adhesi, proliferasi dan diferensiasi hASC juga dievaluasi.

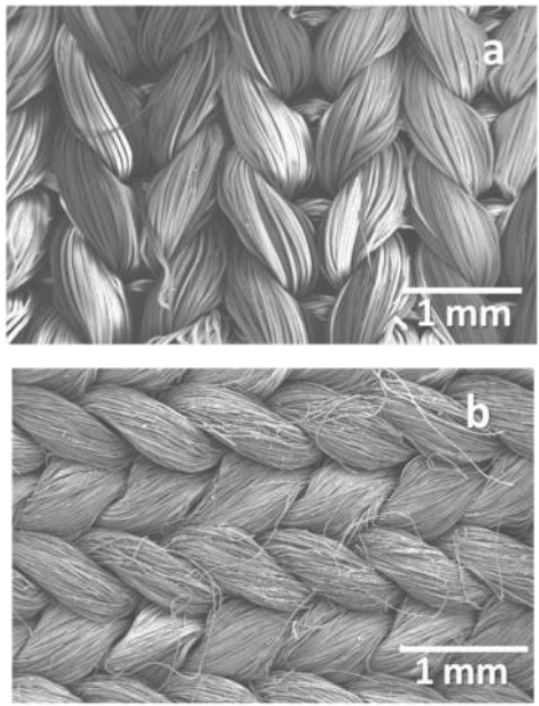

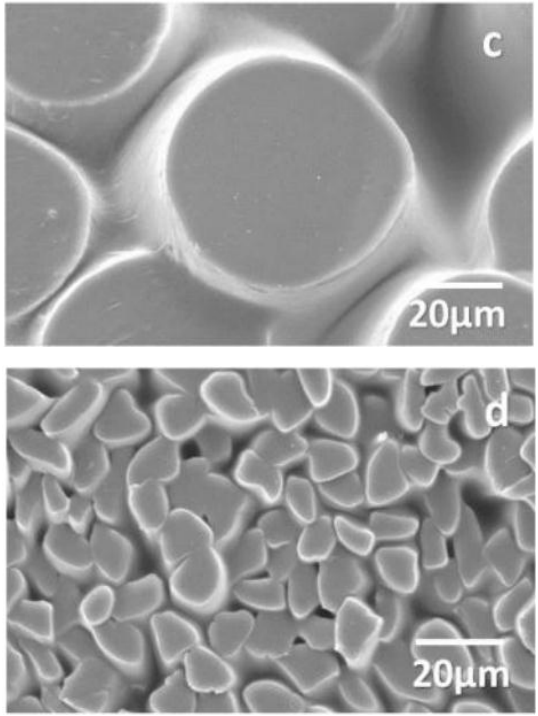

Gambar-11 Kenampakan struktur mikrograf SEM dari a) PBF dan b) SF, c) dan d) masing-masing merupakan penampang lintang kain tersebut

Ribeiro dkk (2013) telah membuat sebuah material Granulated polybutylene suksinat (PBS), material tersebut diperoleh dari Showa Highpolymer Co. Ltd., Tokyo, Jepang. Sutra yang berasal dari ulat sutra Bombyx Mori dalam bentuk kepompong pun diperoleh dan dipintal menjadi benang oleh sericulture APPA-CDM (Asosiasi Orangtua dan Sahabat Remaja Orang Cacat Portugis), Portugal.

Serat PBS yang terdiri atas 36 filamen diproses dan dioptimalkan dalam ekstruder multikomponen (Hills, Inc., West Melbourne, FL, USA). *The Melt Flow Rate* (MFR) ditentukan untuk mendapatkan pemrosesan yang memadai. Peralatan Peleburan Modular dengan pemotong otomatis digunakan sesuai standar ASTM D1238. Pengujian dilakukan pada suhu 190ºC dengan penerapan gaya 2,160 kg. Hasilnya memungkinkan untuk menentukan parameter optimal untuk proses ekstrusi: pengeringan awal dua jam pada suhu 60ºC; profil termal: 120-130ºC; rasio imbang: 2; kecepatan: 300 sampai 600 m / menit.

Secara umum, kedua konstruksi (PBS dan SF) menyajikan porositas yang relatif tinggi dan interkonektivitas 100%, yang berarti bahwa semua pori-pori saling terkait. Meskipun kerapatan linier serat dan parameter pemrosesan yang sama digunakan untuk membangun struktur rajutan, matriks SF menghasilkan porositas yang secara signifikan lebih rendah (68,4 ± 3,7 untuk SF dan 78,4 ± 2,3 untuk PBS) dan ukuran pori dibandingkan dengan PBS (54,5 ± 9,4 untuk SF dan 72,4 ± 13,0 untuk PBS). Perbedaan ini dapat dibenarkan dengan perbedaan ketebalan filamen masing-masing. Ukuran pori rata-rata dari matriks SF dan PBS cocok untuk aplikasi di TE [Hutmatcher dkk, 2007]. Saat mempertimbangkan regenerasi kulit dan penyembuhan luka, misalnya, sebuah studi oleh O'Brien dkk. [O'Brien dkk, 2005] telah menunjukkan bahwa kisaran kritis dari ukuran pori perancah berbasis kolagen adalah antara 20 dan 120 µm untuk memungkinkan aktivitas seluler yang optimal dan sekaligus menghalangi kontraksi luka. Dalam kasus rekayasa jaringan tulang ukuran pori minimal telah dipertimbangkan sekitar 75-100 µm [Hutmatcher dkk, 2007]. Hal ini disebabkan oleh ukuran sel dan migrasi, dan kebutuhan transportasi nutrisi / oksigen. Ukuran pori ini juga dikaitkan dengan kemampuan vaskularisasi. Dengan teknologi merajut sekarang dimungkinkan untuk menyesuaikan dengan ketepatan *interloop* ruang dari matriks tekstil sehingga porositas akhir (dan ukuran pori) dapat disesuaikan sesuai dengan kebutuhan spesifik.

| Surface treatment | XPS ratio O/C (a.u) | Contact angle (°) | Roughness (Ra) | Fibroblasts cell line (L929) after 24 hours | | Fibroblasts cell line (L929) after 14 days | |
|---|---|---|---|---|---|---|---|
| Control | 0.23 | 105.7 ± 1.8 | 21.82 ± 3.23 | 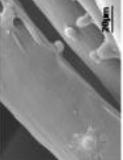 | 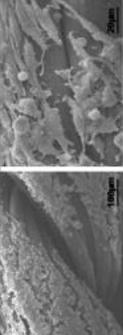 | | |
| NaOH | 0.36 | 61.3 ± 5.8 | 111.78 ± 2.48 | 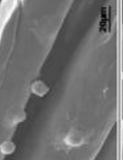 | 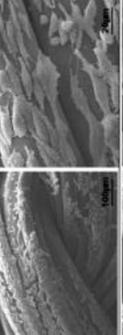 | | |
| UV/O3 | 0.31 | 93.7 ± 3.4 | 36.34 ± 7.02 | 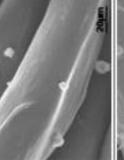 | 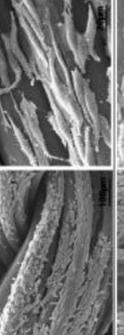 | | |
| Plasma/VSA | 0.34 | 116.6 ± 5.5 | 32.37 ± 10.06 | 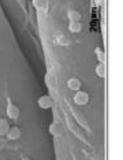 | 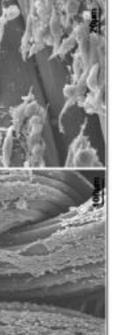 | | |

Tabel-6 Hasil teknik *culturing* kain rajut PBS dengan berbagai *treatment* permukaan

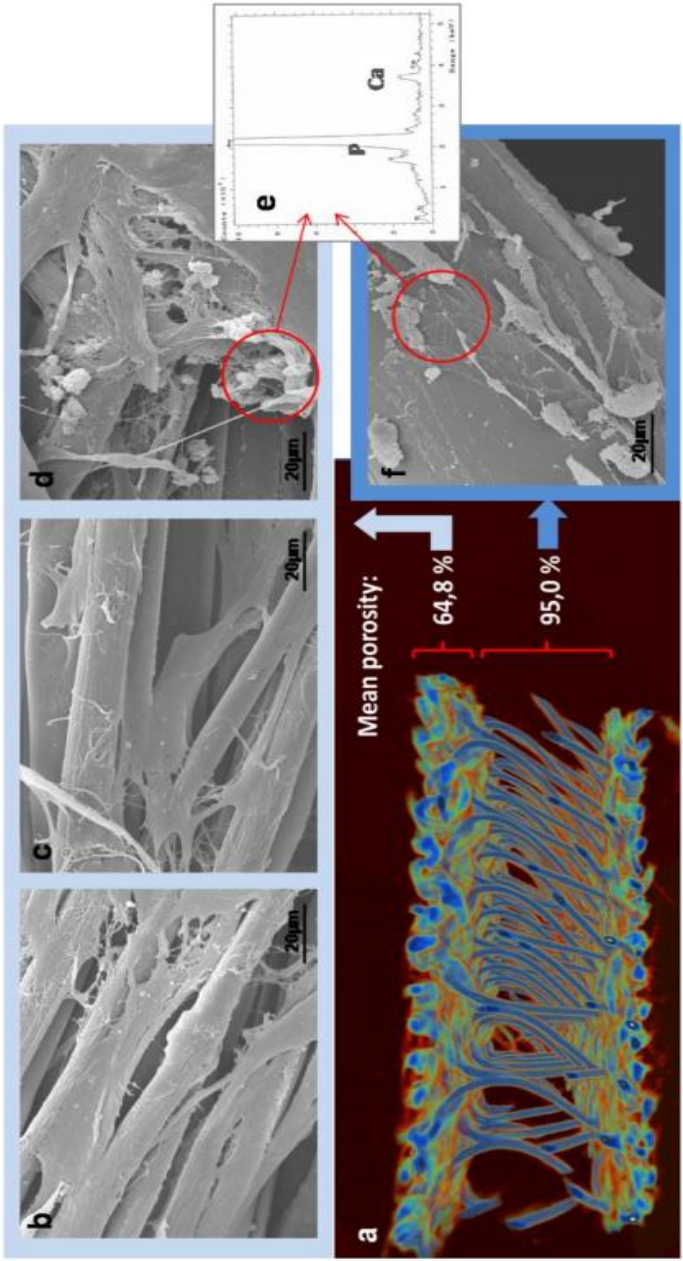

Gambar-12 Struktur kain spacer 3D fibroin sutra, a) rekonstruksi $\mu$-CT 3D, b) c) f) dan d) citra mikrograf SEM yang menunjukan kenampakan hASC's setelah proses kultur selama 28 hari, e) Spektrum EDS kain setelah proses kultur 28 hari

Benang sutera alam diolah menjadi struktur spacer 3D dengan teknologi rajutan lusi. Arsitektur 3D kompleks yang diperoleh terdiri dari dua lapisan rajutan sutra yang dirakit dan ditempatkan oleh monofilamen polietilena tereftalat (PET) untuk meningkatkan dimensi dan kekokohan dari perancah. Gambar-12 menunjukan morfologi 3D dari perancah yang diperoleh dan porositas yang dihitung dengan analisis µ-CT (a) dan mikrograf SEM yang menunjukkan morfologi dan kelekatan hASC pada serat (b-d).

Sel Stem yang berasal dari Adiposa Manusia (hASCs) menjadi kemungkinan munculnya terapi penggantian jaringan. Karena potensi diferensiasi osteogenik tersebut, isolasi mudah, serta kemungkinan perluasan dan proliferasi in vitro, material tersebut telah menunjukkan prospek yang menjanjikan dalam regenerasi tulang. Analisis SEM (Gambar-12) menunjukkan bahwa pada titik awal hASC melekat pada perancah SF yang menyajikan morfologi fibroblastik khasnya, dengan tingkat penyebaran yang lebih tinggi di atas permukaan konstruksi (Gambar-12b). Setelah 14 hari dalam kultur, kolonisasi sel yang luas dapat diamati (Gambar-12c). Setelah 28 hari matriks mineral diendapkan di atas permukaan tekstil (Gambar-12d), seperti yang dikonfirmasi oleh puncak kalsium dan fosfor yang dideteksi oleh analisis spektroskopi sinar-X Energi-dispersif (Gambar-12e). Ketika menganalisis penampang perancah, ada kemungkinan untuk mengamati bahwa sel-sel mampu menembus ke dalam perancah dan menjajah monofilmon PET (Gambar-12f) dengan bukti besar mineralisasi matriks ekstraselular (ECM).

Analisis *mean porosity* diperoleh dengan menerapkan metode pengolahan citra. Citra yang ditangkap dengan menggunakan perangkat SEM kemudian dianalisis dengan tujuan untuk menemukan sifat *porosity* kain terkait. Berdasarkan hasil pengamatan, bahwa diperoleh sekitar 65% rongga pada bagian kain dan 95% pada bagian *space* kain. Hal tersebut menunjukan bahwa kain hasil regenerasi tersebut memiliki karakteristik pori yang cukup untuk dapat digunakan sebagai material hASCs.

# 5. EVALUASI STRUKTUR KAIN RAJUT 3D DENGAN MENGGUNAKAN ANALISIS CITRA

Kenyamanan pakaian adalah salah satu perhatian utama produsen tekstil dan garmen saat ini. Hal ini didasarkan pada kepekaan manusia terhadap bahan pakaian yang ditentukan oleh parameter termal, fisiologis dan mekanik kain terkait. Untuk kain yang bersentuhan langsung dengan kulit, sifat sentuhan sangat penting dalam kaitannya dengan kenyamanan pakaian. Bagian penting dari sifat kenyamanan mekanik adalah kekasaran permukaan kain. Di bidang tekstil, kekasaran biasanya dapat ditangani melalui proses finishing khusus. Kekasaran akan mempengaruhi tampilan dan pegangan yang nyaman [Militky dan Mazal, 2007]. Kekasaran permukaan secara konvensional diukur dengan menggunkan metode profil stylus, dimana profil permukaan yang diukur sebagai fungsi *surface height variation trace* [Greenwood dkk, 1984, Kawabata, 1980]. Selain metode konvensional, adapula metode modern yang dapat mengukur kekasaran permukaan kain didasarkan pada pengolahan citra gambar dari permukaan kain. Kekasaran permukaan tekstil polos sebenarnya telah diidentifikasi pula dengan metode gesekan, *contact blade*, *lateral air flow*, *thickness meter* atau dengan suatu metode penilaian subjektif [Militky, 2007, Greenwood, 1984, Kawabata, 1980, Ajayi, 1994 dan Stockbridge, 1957].

Karakteristik standar profil permukaan kain sebenarnya didasarkan pada variabilitas relatif sebagai suatu fungsi koefisien variasi (sama halnya seperti evaluasi ketidakrataan benang). Parameter standar yang menggambarkan kekasaran permukaan teknis biasanya mengacu pada standar ISO 4287. Untuk karakterisasi kekasaran permukaan tekstil, biasanya dijabarkan sebagai bentuk mean absolut deviation (SMD).

Banyak peneliti telah meneliti sifat permukaan kain, termasuk sifat gesekannya [Kenins, 1994, Kim, 1992, Peykamian, 1999, Vitra, 1997], dan telah mengembangkan instrumen untuk mengukur sifat permukaan kain, seperti Sistem Evaluasi Kawabata (KES-FB) [Zurek, 1985]. Metode ini menentukan ekspresi subjektif dengan mengukur sifat fisik dan mekanik dari kain tekstil dan mengekspresikan pegangan kain tekstil yang diobjekkan dengan menganalisis korelasi timbal balik. Sistem KES-F dapat melakukan pengukuran kekasaran geometrik dan koefisien gesekan (μ) kain secara bersamaan. Elemen penginderaan pada sistem Kawabata terdiri dari batang logam yang dilengkapi dengan kawat tipis dalam bentuk U di ujung logamnya [Savvas, 2004]. Variasi struktur profil permukaan dapat diukur secara kuantitatif, dan korelasi antara sifat fisik dan sensasi subjektif memiliki hubungan yang cukup erat. Namun, metode KES memakan waktu, dan terjemahan data yang diukur dinilai sulit untuk menggambarkan citra kain secara 3D. Oleh karena itu, metode sederhana telah dirancang, seperti metode ekstraksi [Kim, 1992, Peykamian, 1999, Yoon, 1984 dan Hasani, 2009] dan metode kereta luncur [Ajayi, 1992, Carr, 1994 dan Yoon, 1984]. Sifat permukaan kain telah dipelajari dalam kaitannya dengan jenis, sifat tidak mudah terbakar dan kadar air pada suatu kain [Kim, 1992]. Metode kereta luncur [Ajayi, 1992 dan Carr, 1994] telah digunakan untuk mengukur kekuatan ekstensional untuk meregangkan dua lembar kain. Ujung jari manusia atau bagian belakang tangan sebenarnya juga telah digunakan untuk menemukan kekasaran yang dinilai lebih realistis [Kim, 1992]. Karena pengukuran tipe kontak lebih mudah dipengaruhi oleh kondisi lingkungan, seperti kelembaban, serta membutuhkan lebih banyak waktu daripada metode non kontak, maka metode ini tidak sesuai untuk sistem *real time* [Militky, 2007]. Metode modern didasarkan pada pengolahan citra gambar permukaan atau gambar kain. Kekasaran permukaan tekstil polos sebenarnya telah diidentifikasi pula dengan metode gesekan, *contact blade*, *lateral air flow*, *thickness meter* atau dengan suatu metode penilaian subjektif [Militky, 2007, Greenwood, 1984, Kawabata, 1980, Ajayi, 1994 dan Stockbridge, 1957].

Semnani dkk (2011) telah melakukan penelitian mengenai kekasaran dengan menggunakan *scanner* beresolusi tinggi untuk mengevaluasi kekasaran permukaan kain rajut. Analisis citra digunakan untuk ekstraksi profil permukaan. Hasil yang diperoleh dari analisis citra dibandingkan dengan nilai SMD yang diukur dengan metode Kawabata. Selain itu, pengaruh parameter serat, benang dan kain terhadap kekasaran kain (SMD) juga telah dianalisis.

Tabel-7 Jenis dan spesifikasi kain rajut yang digunakan oleh Semnani dkk (2011)

| Code | Yarn properties | Fabric structure | Loop/cm² | Processing stage |
|---|---|---|---|---|
| A01 | Cotton, Ring, Ne25 | Double cross tuck | 208 | Bleaching |
| A02 | | Double cross miss | 165 | |
| A03 | | Plain single jersey | 352 | |
| A04 | | | 282 | |
| A05 | | | 195 | |
| A06 | | Plain rib | 92 | Dyeing |
| A07 | Viscose, Ring, Ne25 | | 60 | |
| A08 | Cotton, Ring, Ne30 | Interlock | 268 | |
| A09 | Cotton, Ring, Ne24 | | 254 | |
| A10 | Cotton, Compact, Ne25 | Plain single jersey | 314 | Bleaching |
| A11 | Cotton, Ring, Ne25 | | 312 | |
| A12 | Cotton, Open-end, Ne25 | | 314 | |
| A13 | Cotton, Ring, Ne25 | | 310 | Bleaching & Softening (2%) |
| A14 | | | 315 | Dyeing |
| A15 | | | 310 | Bleaching & Softening (4%) |

Tabel-8 Spesifikasi proses penyempurnaan yang dilakukan pada kain

| Processing stages | Description |
|---|---|
| Bleaching | at 80 °C for 30 minute with Hydrogen peroxide (1.5%) and rinsed |
| Dyeing | with a reactive dye (Bezative orange S-RL 150 ) at 60 °C and rinsed |
| Softening | 2 and 4% Tubingal KRE at 40 °C for 20 minute |

Lima belas jenis kain rajutan digunakan dalam penelitian ini, spesifikasi kain-kain tersebut ditunjukkan pada Tabel-7. Kain rajutan ini diproduksi pada mesin rajut bundar dengan struktur serat, benang dan kain yang berbeda. Kemudian kain itu dilakukan di tahap *finishing* yang berbeda. Spesifikasi penyempurnaan ditunjukkan pada Tabel-8. Nilai SMD dari sampel kain diukur dengan instrumen KES-FB4. Untuk setiap sampel, setiap pengukuran dilakukan dua kali untuk tiga sampel terpisah yang dipotong dari titik tengah kain rajutan, dan enam nilai yang dihasilkan dirata-ratakan. Sampel berukuran standar 200 mm x 200 mm diuji dalam arah wale dan course. Karena anisotropi adalah pertimbangan pada kain rajutan, kekasaran permukaan diukur ke arah *wale* dan *course*. Rata-rata pengukuran *wale* dan *course* dihitung untuk analisis lebih lanjut. Pada tahap persiapan contoh uji, pengkondisian kain dilakukan pada kondisi atmosfir standar: suhu 20 ± 2°C dan kelembaban relatif 65 ± 2%.

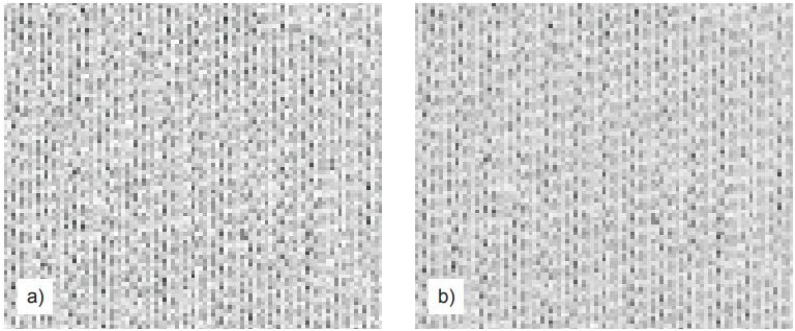

Gambar-13 Kenampakan citra digital kain rajut, a) gambar asli, b) citra hasil proses *filtering* dengan menggunakan metode pengolahan citra digital

Profil permukaan kain rajutan dipindai menggunakan *scanner* resolusi tinggi. Area gambar yang ditangkap dibatasi sampai ukuran 10 cm x 10 cm. Kekasaran kain rajutan tergantung pada banyak faktor yang dapat dikelompokkan menjadi kelompok faktor material dan struktural. Efek material pada fitur permukaan terutama disebabkan oleh jenis benang, yang meliputi jumlah benang, formasi serat, *twist* benang dan migrasi serat. Efek struktural dapat dipengaruhi oleh kerapatan *stitch*, panjang *loop* dan ketebalan kain.

Permukaan kain yang ideal tidak boleh dianggap sebagai permukaan yang datar karena permukaan nyata kain tidak akan benar-benar rata seluruhnya. Dalam metode KES, jarum *probe* pada perangkat tersebut bisa merasakan permukaan kain. Untuk mengevaluasi kekasaran permukaan, diperlukan perbandingan atau acuan agar permukaan kain bisa dibandingkan. Sebenarnya, referensi ini tidak lain adalah lapisan yang dapat dikatakan sebagai kekasaran yang paling nyaman untuk sentuhan manusia, profil permukaan ini disebut sebagai permukaan ideal. Kain dengan permukaan ideal tersebut dapat dilihat pada Gambar-14.

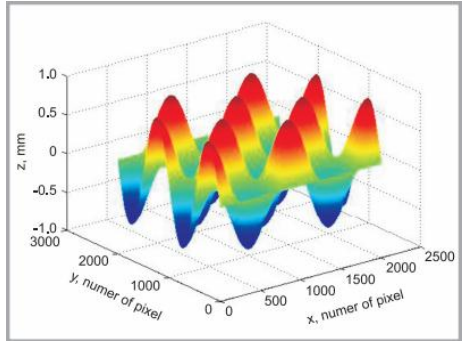

Gambar-14 Pemodelan struktur permukaan ideal kain rajut

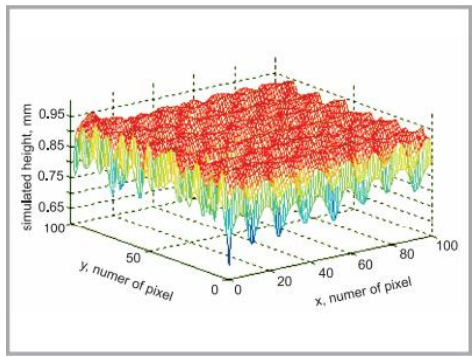

Gambar-15 Hasil pemindaian profil kain menggunakan prinsip pengolah citra

Pertama, kain rajut dipindai dengan resolusi 600 DPI menggunakan *scanner*, dan lapisan hitam disisipkan pada *scanner* sebagai latar belakang gambar. Gambar yang diperoleh diubah menjadi gambar skala abu-abu, dan kemudian dilakukan filter Wiener dan Gaussian pada gambar untuk mengurangi *noise* di dalamnya. Dalam gambar ini, zona terang menunjukkan bagian padat dari lapisan kain, dan zona gelap menunjukkan bagian lapisan yang jarang. Gambar-13 menunjukkan contoh gambar pindaian asli dan gambar hasil proses.

Setiap piksel dalam gambar ini dipetakan sebagai titik profil permukaan kain. Oleh karena itu nilai setiap elemen dalam profil ini adalah simulasi tinggi titik pada gambar tersebut. Daerah paling terang pada gambar dipetakan ke titik tertinggi dalam profil. Akibatnya, pada gambar skala abu-abu, pixel yang sama dengan 255, menggambarkan titik tertinggi di lapisan, yang tingginya merupkan ketebalan kain yang dicapai dari percobaan. Tinggi ini dipilih untuk membandingkan permukaan contoh uji dengan permukaan dari hasil simulasi, dan juga karena semua kekasaran permukaan spesimen harus dibandingkan dengan koefisien gesekan specimen, maka penting agar semua gambar ini dibuat dalam format individual. Format individual ini adalah satu yang juga sesuai dengan permukaan simulasi yang ideal secara dimensional. Hal ini akan menyebabkan profil yang amplitudonya tingginya berbeda dari 0 sampai ketebalan kain dikatakan sama atau sesuai dengan prediksi pemodelan. Gambar-14 menunjukkan profil permukaan simulasi yang dihasilkan oleh algoritma simulasi.

Terdapat lima parameter kriteria profil permukaan dievaluasi untuk menilai kekasaran permukaan. Kriteria ini adalah sebagai berikut: (1) n: jumlah puncak pada permukaan kain, di mana puncak didefinisikan sebagai titik di mana tingginya lebih tinggi daripada titik sekitarnya, dan memiliki ketinggian yang sama dengan beberapa puncak lainnya. ; (2) v: varians dari jarak puncak ke titik asal; (3) s: volume profil simulasi dari gambar. (4) a: rasio varians dengan mean tingkat skala abu-abu gambar, dan (5) g: varians dari tingkat puncak skala abu-abu pada profil permukaan kain. Pada penelitian Semnani (2011), kelima kriteria ini digunakan untuk mendefinisikan faktor ketajaman permukaan, yang ditunjukkan sebagai Kt. Dengan menggunakan lima kriteria ini, kita dapat mengevaluasi faktor kekasaran permukaan. Sebenarnya, nilai setiap elemen yang mewakili nilai tinggi dalam profil simulasi digunakan untuk menilai kekasaran. Gagasan menggunakan nilai tinggi untuk menilai kekasaran berasal dari KES.

Kelima kriteria tersebut diukur dari gambar, dan selanjutnya kriteria ini dibandingkan dengan simulasi permukaan ideal. Untuk menghitung data yang dievaluasi dari gambar dan menyimpulkannya dalam satu kriteria, faktor Kt dihitung untuk setiap gambar dengan menggunakan data yang dibandingkan, dihitung dengan menggunakan persamaan yang diusulkan pada Tabel-9. Untuk setiap sampel, setiap pengukuran dibuat untuk lima sampel terpisah, yang dipotong dari titik tengah kain, dan lima nilai yang dihasilkan langsung dirata-ratakan.

Tabel- 9 Kriteria faktor kekasaran kain

| Parameter | Formula | Descriptions |
|---|---|---|
| $K_1$ | $\dfrac{n - ni}{n}$ | n & ni are the number of peak points in an ideal and real profile |
| $K_2$ | $\dfrac{v - vi}{v}$ | v & vi are the Variance of the distance vector in an ideal and real profile |
| $K_3$ | $\dfrac{s - si}{s}$ | s & si arethe Integral of the surface under peaks for an ideal and real profile |
| $K_4$ | $\dfrac{a - ai}{a}$ | a & ai are the Variance of an ideal and real surface |
| $K_5$ | $\dfrac{g - gi}{g}$ | g & gi are cv% of an ideal and real surface |
| $K_t$ | $\dfrac{K1 + K2 + K3 + K4 + K5}{1000}$ | Roughness index achieved by image processing |

Tabel-10 menunjukan nilai kekasaran yang diperoleh dari dua sampel yang berbeda, yakni sampel kain A01 dan sampel kain A08. Berdasarkan hasil pengamatan tersebut, bahwa kain dengan struktur yang berbeda akan menghasilkan kekasaran permukaan yang berbeda. Pada Tabel-11 menunjukan perbandingan hasil pengukuran kekasaran kain dengan menggunakan kedua metode, yaitu metode Kawabata dan metode pengolah citra digital Semnani dkk (2011)

Tabel-10 Nilai kekasaran yang diperoleh dari kain jenis A01 dan A08

| Picture no. | Roughness value for Sample A01 | Roughness value for Sample A08 |
|---|---|---|
| 1 | 0.1155 | 0.0457 |
| 2 | 0.1128 | 0.0472 |
| 3 | 0.1160 | 0.0410 |
| 4 | 0.0858 | 0.0427 |
| 5 | 0.1100 | 0.0406 |
| 6 | 0.1402 | 0.0386 |
| 7 | 0.1159 | 0.0349 |
| 8 | 0.1107 | 0.0475 |
| 9 | 0.0998 | 0.0424 |
| 10 | 0.1285 | 0.0420 |
| 11 | 0.1241 | 0.0340 |
| 12 | 0.1014 | 0.0514 |
| 13 | 0.1358 | 0.0482 |
| 14 | 0.1298 | 0.0438 |
| 15 | 0.1149 | 0.0438 |
| 16 | 0.1145 | 0.0446 |
| 17 | 0.1163 | 0.0389 |
| 18 | 0.1144 | 0.0407 |
| 19 | 0.1196 | 0.0538 |
| 20 | 0.0926 | 0.0450 |

Tabel-11 Perbandingan nilai kekasaran yang diperoleh dengan metode Kawabata dan pengolah citra digital

| Fabric code | SMD (Kawabata method), μm | | Roughness index ($K_t$) (Image processing) | |
|---|---|---|---|---|
| | Average | SD | Average | SD |
| A01 | 18.74 | 0.1 | 0.0457 | 0.80 |
| A02 | 12.52 | 0.9 | 0.0694 | 0.75 |
| A03 | 5.51 | 0.5 | 0.1093 | 0.90 |
| A04 | 7.98 | 1.1 | 0.1042 | 1.00 |
| A05 | 8.96 | 0.8 | 0.0998 | 1.20 |
| A06 | 5.37 | 1.0 | 0.1301 | 0.50 |
| A07 | 3.57 | 1.0 | 0.1402 | 1.00 |
| A08 | 5.84 | 0.9 | 0.1155 | 0.65 |
| A09 | 7.12 | 0.4 | 0.1146 | 0.90 |
| A10 | 7.34 | 0.55 | 0.1023 | 0.70 |
| A11 | 6.35 | 0.8 | 0.1260 | 1.30 |
| A12 | 11.52 | 0.8 | 0.0921 | 1.00 |
| A13 | 5.59 | 1.1 | 0.1324 | 0.95 |
| A14 | 6.77 | 0.9 | 0.1209 | 0.85 |
| A15 | 5.40 | 0.85 | 0.1367 | 0.78 |

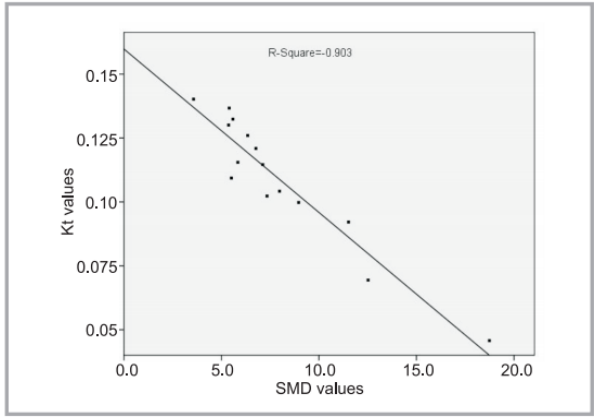

Gambar-16 Hubungan antara hasil pengukuran kekasaran kain dengan menggunakan metode Kawabata (sumbu x) dan metode pengolahan citra digital (sumbu y)

Untuk mengetahui hubungan antara indeks kekasaran yang diukur dengan pengolahan citra dan nilai SMD yang diukur dengan metode Kawabata, Semnani (2011) melakukan analisis regresi dengan menggunakan perangkat lunak statistik SPSS. Gambar-16 menunjukkan plot regresi dari analisis tersebut. Koefisien korelasi hasil perhitungan adalah 0,903, menunjukkan korelasi yang baik antara nilai kekasaran kain yang diukur dengan dua metode yang berbeda. Selain itu, koefisien korelasi negatif menunjukkan bahwa nilai kekasaran yang diukur dengan perubahan Kawabata berbanding terbalik dengan yang diukur dengan metode pengolahan citra.